\documentclass{ws-mpla}

\begin{document}

\markboth{TORU KOJO and DAISUKE JIDO}
{LIGHT SCALAR NONETS IN POLE-DOMINATED QCD SUM RULES}

%%%%%%%%%%%%%%%%%%%%% Publisher's Area please ignore %%%%%%%%%%%%%%
\catchline{}{}{}{}{}
%%%%%%%%%%%%%%%%%%%%%%%%%%%%%%%%%%%%%%%%%%%%%%%%%%%%%%%%%%%%%%%%%%%

\title{SCALAR NONETS IN POLE-DOMINATED QCD SUM RULES}

\author{\footnotesize TORU KOJO }

\address{Department of Physics, Kyoto University, Kyoto 606-8502, Japan\\
torujj@ruby.scphys.kyoto-u.ac.jp}

\author{DAISUKE JIDO}

\address{Yukawa Institute for Theoretical Physics, Kyoto University,
          Kyoto 606-8502, Japan}

\maketitle

\pub{Received (Day February 2008)}{Revised (Day Month Year)}

\begin{abstract}
The light scalar nonets
are studied using the QCD sum rules for the tetraquark operators. 
The operator product expansion for the correlators is calculated
up to dimension 12 and this enables us to perform 
analyses retaining sufficient pole-dominance.
To classify the light scalar nonets,
we investigate the dependence on current quark mass and flavor dynamics.
Especially, to examine the latter,
we study separately SU(3) singlet and octet states,
and show that the number of annihilation diagrams is largely responsible
for their differences,
which is also the case even after the inclusion of the finite quark mass.
Our results support the tetraquark picture for isosinglets,
while that for octets is not conclusive yet.

\end{abstract}

\ccode{PACS Nos.: 12.39.Mk, 11.55.Hx, 11.30.Rd}

%%%%%%%%%%%%%%%%%%%%%%%%%%%%%%%%%%%%%%%%%%%%%%%%%%%%%%%%%%%%%%%%%%%%%%%%%%
\section{Light scalar nonets as tetraquarks}
The structure of scalar mesons is a long-standing
problem in hadron physics\cite{review}.
In contrast to the other hadrons, two flavor nonets appear around 1 GeV 
in the scalar meson spectra. 
Especially, the lighter scalar nonets ($\sigma(600),\kappa(800),a_0(980),f_0(980)$)
are candidates for tetraquark states
since the $qq\bar q \bar q$ assignments 
for these mesons naturally explain their mass ordering and decay patterns
in contrast to the $q\bar q$ assignments.
These explanations are qualitatively favorable, 
but there still remain several questions to be answered by the quantitative studies:
i) Can $qq\bar{q}\bar{q}$ configurations provide light masses below
1 GeV despite of large number of quarks?
If possible, which effects are responsible for the mass reduction?
ii) Do all states in nonets have large tetraquark components?
iii) How large are current quark mass effects on the mass splitting?
We attempt to answer these questions with nonperturbative treatments
of correlators for interpolating fields of light scalar nonets. 

Especially, i) and ii) are deeply related to possible intermediate states depending on 
flavor structure of the nonets rather than current quark mass effects.
To see this, it is convenient to consider the SU(3) singlet 
(${\cal S}$) and octet (${\cal O}_i(i$=1$\sim$8))
states in the SU(3) chiral limit.
Using diquark bases, 
$U$=$(\bar{d}\bar{s})$, $D$=$(\bar{s}\bar{u})$, $S$=$(\bar{u}\bar{d})$,
these states are described as
${\cal S} = ( U\bar{U} + D\bar{D} + S \bar{S})/\sqrt{3},\ 
{\cal O}_{1} = U \bar{D},\ 
{\cal O}_{2} = ( U\bar{U} + D\bar{D} -2 S \bar{S})/\sqrt{6},...
$
The isodoublet $\kappa$ and isovector $a_0$ belong to 
purely the octet because of the nonzero isospin, while 
the isoscalar $\sigma$ and $f_0$ can be composed of the
mixture of the singlet and octet sates in the real world where
the flavor SU(3) symmetry is broken by the quark masses.
Thus the ideal mixing is expected to be realized as,
$\sigma \sim  \bar{S} S 
 = \sqrt{1/3} {\cal S}
- \sqrt{ 2/3 } {\cal O}_{2},
\ 
f_0 \sim \sqrt{1/2} [ \bar{D}D + \bar{U}U ]
 = \sqrt{ 2/3 } {\cal S}
   + \sqrt{ 1/3 } {\cal O}_{2}. 
\label{f0}
$

\begin{figure}[floatfix]
  \begin{center}
   \resizebox{50mm}{24mm}{\includegraphics{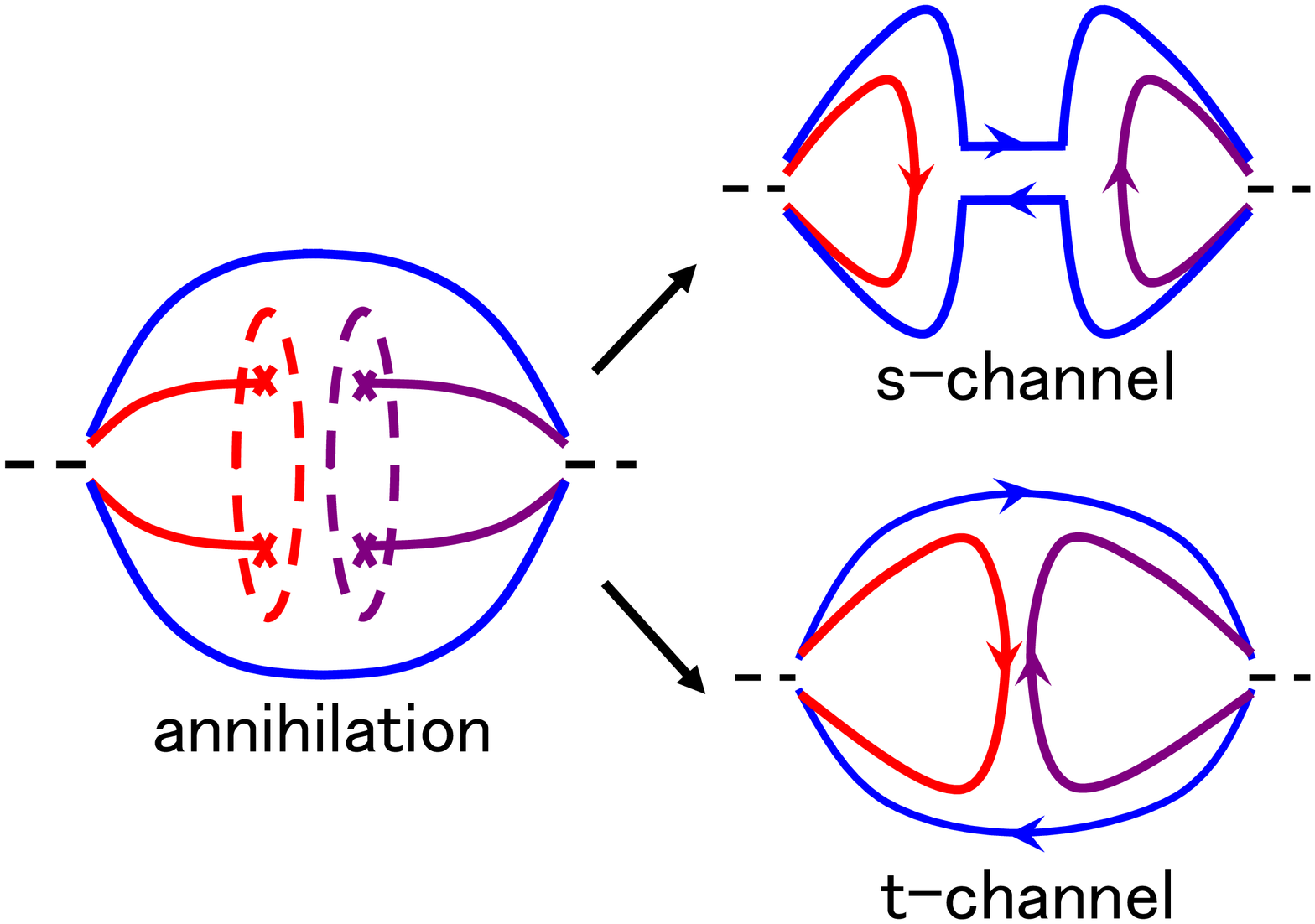}}
   \caption{ The two ways of interpretation
   of the annihilation diagrams.
   The dashed circle represents a pair of quarks
   forming the condensates.
   By deforming the quark line,
   we can interpret the annihilation diagrams as 
   either 2q like s-channel propagation 
   or 4q like t-channel propagation with
   exchange of mesonic resonances.
   }
   \label{anni}
   \vspace{-0.5cm}
  \end{center}
\end{figure}

Here let us see the difference between singlet and octet states.
The difference emerges from the number of the $q\bar q$ annihilation
diagrams \cite{KJ} in which some quark lines disconnected
between the space-time points x and 0 like Fig.\ref{anni}.
For example, the correlator included in singlet case,
$[\bar U U(x) \bar U U (0)]=[ ds\bar{d}\bar{s}(x) ] 
[\bar{d}\bar{s}ds(0)]$, 
has larger number of annihilation diagrams than 
that in octet case,
$ [ \bar{U}D(x) \big] \big[ \bar{U}D(0) ]
 = [ ds\bar{s}\bar{u}(x) \big] \big[ \bar{d}\bar{s}su(0) ]$.
We can verify that the ratio of number of annihilation diagrams is
${\cal O}:\sigma:f_0:{\cal S}=1:2:3:4$.

We note that the annihilation diagrams 
do not always represent the 2q mesonic propagations.
As shown in Fig.\ref{anni},
annihilation diagrams can be interpreted as either
s-channel and t-channel processes.
Especially, we can interpret the latter as
diquark-antidiquark correlation
which was conjectured to largely reduce masses
of the tetraquark states\cite{Jaffe}.
We will see that the annihilation diagrams 
play key roles in the Borel analyses.

In later analyses, we will use the current
$J(\theta) = \cos\theta \, J_P + \sin \theta\,  J_S$ with 
$J_P =\epsilon^{abc} \epsilon^{dec} 
 [u_a^T C d_b][\bar{u}_d C \bar{d}_e^T]$
and 
$J_S = \epsilon^{abc} \epsilon^{dec}
 [u_a^T C\gamma_5 d_b][\bar{u}_d \gamma_5 C \bar{d}_e^T]$.
$\theta$ will be chosen to achieve
the pole-dominance and the small threshold dependence
with better degree.

\section{QCD Sum Rules and Borel analyses}
We analyze tetraquark correlators 
using the QCD sum rules (QSR)\cite{Shifman}
which relate the nonperturbative aspects of QCD
to the hadronic properties through the dispersion relation for 
the correlator $\Pi$.
The Borel transformed QSRs with using the quark-hadron duality
above $s_{th}(= E_{th}^2)$ are described as follows:
\begin{eqnarray} 
\int_0^{s_{th}} \!\! ds\ e^{-s/M^2}\frac{1}{\pi} {\rm Im}\Pi^{<}(s)
= L_M \Pi^{ope}(-Q^2) 
- \int_{s_{th}}^{\infty} \!\! ds\ e^{-s/M^2}\frac{1}{\pi} {\rm Im}\Pi^{ope}(s),
\label{eq:disp}
\end{eqnarray}
where RHS includes the correlator $\Pi^{ope}$ calculated by operator product
expansion (OPE) and the hadronic parameters in LHS are evaluated as
outputs.

Using Eq.(\ref{eq:disp}), the {\it effective} mass and residue are evaluated by
\begin{eqnarray}
m_{ {\rm eff} }^2(M^2) \equiv \frac{ \int_0^{s_{th}} \!ds e^{-\frac{s}{M^2} }s{\rm Im} \Pi^< (s) }
 { \int_0^{s_{th}} \!ds e^{-\frac{s}{M^2} }{\rm Im} \Pi^< (s) },
 \label{eq:mass}\ \ 
\lambda_{ {\rm eff} }^2(M^2) \equiv e^{ \frac{m_{ {\rm eff} }^2}{M^2} }
\!\! \int_0^{s_{th}} \!ds
  e^{-\frac{s}{M^2} }{\rm Im} \Pi^< (s), 
\end{eqnarray}
where ``effective'' mass (residue) means that $m_{ {\rm eff} }$ is
averaged mass (residue) in the energy region from 0 to $s_{th}$
including the width and background effects.
If integral is well-saturated by single peak with mass $m$,
$m_{ {\rm eff}}$ approaches to $m$ with weak dependence on the value of
$M$,
while if no large and sharp peak exists, 
$m_{ {\rm eff}}$ shows large dependence on $M$.
The same argument is valid for the residue.
To estimate these quantities of low energy excitations with good accuracy and
small ambiguities,
we need to treat sum rules in the appropriate $M^2$ region
where low energy contributions below $s_{th}$ are large enough
compared to the
contaminations from high energy components which have no relations with 
properties of low-lying resonances\cite{KHJ,KHJ2,Narison}.
Especially, without the sufficient pole contribution,
we are stuck with the {\it pseudo-peak} artifacts
which yield artificial stability of the masses against 
$M^2$ variation\cite{KJ}, which happens often in the QSR for multiquark-hadrons.

For these reasons, we calculate the OPE up to dimension 12
to set the reasonable $M^2$ window where we achieve both
sufficient pole-dominance (pole contribution is larger than 50\% of the total)
and OPE convergence (highest dimension terms are less than 10\% of whole OPE).
We will use QSR within this $M^2$ window.
\begin{figure}[b]
  \begin{center}
   \begin{tabular}{cc}
    \vspace{-0.3cm}
 \hspace{-0.4cm}  \resizebox{63mm}{30mm}{\includegraphics{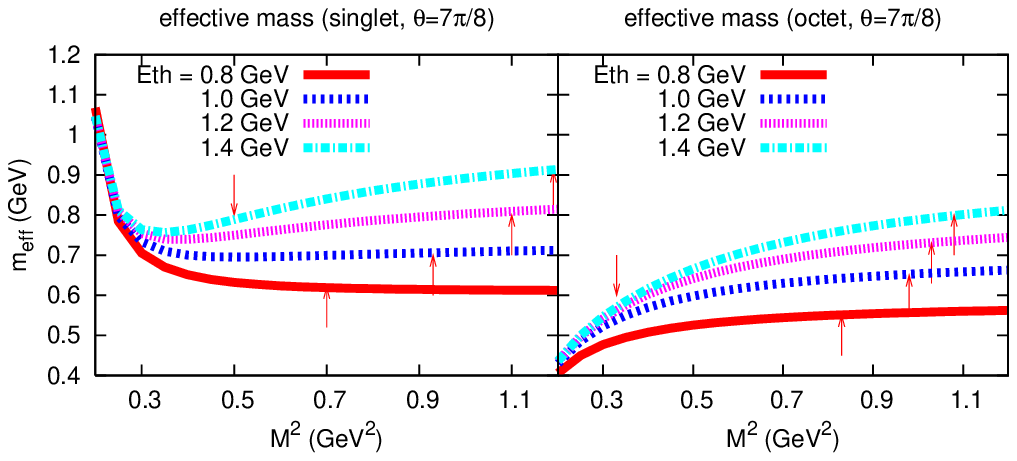}} &
  \hspace{-0.4cm}  \resizebox{63mm}{30mm}{\includegraphics{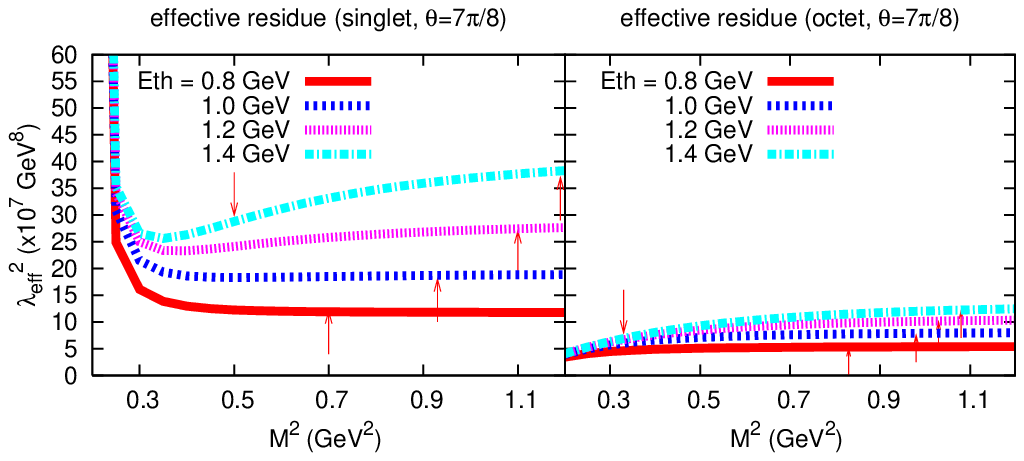}}\\
    \end{tabular}
    \caption{ 
   The effective masses (left) and residua (right) for the singlet and octet states.
   The values of the threshold $E_{th}$ are $0.8,\ 1.0,\ 1.2$ and 1.4 GeV.
   The downward and upward arrows indicate
   the lower and upper bounds of the $M^2$ window,
   respectively. }
    \label{somass}
  \end{center}
\end{figure}

First we show in Fig.\ref{somass} the case of singlet and octet states
($\theta=7\pi/8$)
in the SU(3) chiral limit to see the roles of the annihilation diagrams.
The effective mass for the singlet case is found $700\sim 850$ MeV
in consideration of the possible width effect\cite{KJ}.
For the octet case the effective mass is estimated by $600\sim 750$ MeV,
although the effective mass in the octet channel depends on $M$ fairly
indicating that the signal of the octet resonance is weak and
considerably affected by low energy scattering states.
It should be noted that
the residue of the octet state is smaller than that of the singlet state
by factor $\sim3$.
This means that annihilation diagrams provide more sufficient strength 
for the singlet case than the octet case.
The small strength in the octet case would make
the effective mass fairly depend on $M^2$ value.

\begin{figure}[t]
  \begin{center}
   \resizebox{90mm}{30mm}{\includegraphics{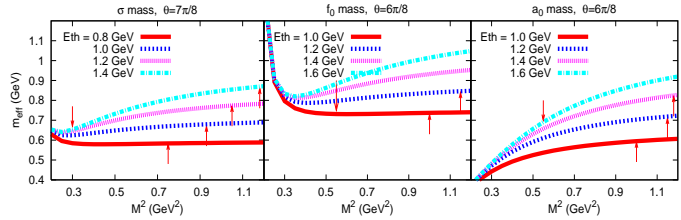}}
    \caption{ 
   The effective masses for the $\sigma,f_0,a_0$ states
   with $\theta=7\pi/8,\ 6\pi/8,\ 6\pi/8$, respectively.
   The downward and upward arrows indicate
   the lower and upper bounds of the $M^2$ window,
   respectively. }
    \label{sfa}
   \vspace{-0.5cm}
  \end{center}
\end{figure}

Including the finite strange quark mass 0.12 GeV, 
we show in Fig.\ref{sfa} are the effective mass plots for
$\sigma(600),\ f_0(980)$, and $a_0(980)$ with 
$\theta = 7\pi/8,\ 6\pi/8,\ 6\pi/8$, respectively.
$\sigma$ and $f_0$ include the singlet component
and show the sufficient pole strength yielding the 
moderate stability in the effective mass plots
around $600\sim 800$ MeV, $750\sim900$ MeV,
respectively.
On the other hand, $a_0$ includes only the octet state
and show the rather large $M$ dependence.
The other octet state, $\kappa$, also shows the same behavior as $a_0$
reflecting its octet nature rather than the quark mass effects.

In conclusion,
we perform the QSR analyses for light scalar nonets
retaining the sufficient pole dominanance.
Our tetraquark correlator analyses show the sufficient
spectral strengths in the region below 1 GeV,
in sharp contrast to the results from two quark meson correlators
yielding typical mass around $\sim$ 1 GeV.
Therefore our results support the tetraquark picture for isosinglets,
while that for octets is still not conclusive
because of rather large $M$ dependence,
which is probably emerged from low energy scattering states.
The origin of these differences are the annihilation diagrams,
and it will be important to deduce more qualitative 
pictures from this fact.

\section*{Acknowledgments}

We thank the organizers and the members of 
Research Center for Nuclear Physics (RCNP) at Osaka University
for hospitality during Chiral07.
%This work is supported by the Grant-in-Aid
%for the 21st Century COE ``Center for Diversity and Universality
%in Physics'' from the MEXT of Japan 
D.J is supported by 
the Grant for Scientific Research (No.\ 18042001).
This research is part of Yukawa International Program for Quark-Hadron Sciences.

\end{document}